\documentclass[pdflatex,sn-mathphys-num]{sn-jnl}

\DeclareUnicodeCharacter{03B1}{\ensuremath{\alpha}} 
\usepackage[utf8]{inputenc}
\usepackage{graphicx}%
\usepackage{multirow}%
\usepackage{amsmath,amssymb,amsfonts}%
\usepackage{amsthm}%
\usepackage{mathrsfs}%
\usepackage[title]{appendix}%
\usepackage{xcolor}%
\usepackage{textcomp}%
\usepackage{manyfoot}%
\usepackage{booktabs}%
\usepackage{algorithm}%
\usepackage{algorithmicx}%
\usepackage{algpseudocode}%
\usepackage{listings}%
\usepackage[version=4]{mhchem}
\usepackage{chemist}
\usepackage{chemfig}
\DeclareUnicodeCharacter{03B1}{\ensuremath{\alpha}} 
\usepackage[utf8]{inputenc}
\usepackage[T1]{fontenc}
\usepackage{charter}
\usepackage[expert]{mathdesign}
\usepackage[utf8]{inputenc}
\usepackage[french,german,english]{babel}
\usepackage{caption}
\usepackage{subcaption}
\usepackage{array}
\usepackage{booktabs} 
\usepackage{float}
\usepackage{sidecap, caption}

\theoremstyle{thmstyleone}%
\theoremstyle{thmstyletwo}%

\theoremstyle{thmstylethree}%

\begin{document}

\title[Article Title]{\textbf{Enhancing Hydrovoltaic Power Generation through Coupled Heat and Light-Driven Surface Charge Dynamics}}


\author{\fnm{Tarique} \sur{Anwar}}
\author{\fnm{Giulia} \sur{Tagliabue}}
\affil{\orgdiv{\text{Laboratory of Nanoscience for Energy Technologies (LNET), STI} }\newline  \orgname{\text{École Polytechnique Fédérale de Lausanne} }, \orgaddress{\postcode{1015}, \state{VD}, \country{Switzerland}}}


\abstract{Harnessing natural evaporation offers a sustainable and untapped pathway for next-generation energy technologies. Here, we present a unified physical and experimental framework for evaporation-driven hydrovoltaic (EDHV) systems that decouples and systematically controls the key interfacial processes underlying electricity generation from ambient heat and sunlight. By introducing an intermediate ion-conducting layer, we spatially and functionally separate the evaporative top interface from the silicon–dielectric nanopillar array at the bottom, enabling independent modulation of evaporation, ion transport, and interfacial chemical equilibrium. This decoupling strategy not only enhances device performance but also allows precise investigation of the mechanisms governing thermal and photo-induced charge generation, thereby affecting ion migration from the bottom electrode and effectively enhancing electricity generation. We develop a predictive equivalent electrical circuit model that captures the coupling between these processes through a transfer capacitance term, which we derive analytically as a function of geometric and material parameters. Our study reveals that capacitive photocharging and thermally modulated surface equilibria—rather than Faradaic or photothermal effects—are the dominant drivers of energy conversion when interfacial environments are adequately engineered. The device achieves a state-of-the-art open-circuit voltage of 1 V and a peak power density of 0.25 W/m$^2$ at a 0.1 M salt concentration. Strategic variation of doping reveals that increasing silicon doping enhances voltage by 28\% and power by 1.6 times, while switching the dielectric shell from TiO$_2$ to Al$_2$O$_3$ boosts voltage (power) by up to 1.9 times (3.6 times). Together, these findings yield insights into advancing EDHV devices and propose broader operational strategies that account for environmental conditions, water salinity, and material engineering to optimize the harnessing of waste heat and sunlight.}

\maketitle

\section{Introduction}
Evaporation, a natural process with an average global energy flux of $(80\ \text{W/m}^2)$ \cite{cavusoglu_potential_2017}, has significant potential for energy harvesting. Hydrovoltaic technology, which generates energy through the direct interaction of materials with water\cite{fang_evaporating_2022, song_conveyor_2024, xue_water-evaporation-induced_2017}, has recently emerged as a promising avenue for sustainable energy generation\cite{zhang_emerging_2018, anwar_salinity-dependent_2024, sun_achieving_2022, shao_boosting_2023}. In particular, evaporation-induced hydrovoltaic devices (EDHV) stand out for their ability to produce continuous (24-hour) electricity output, operating without external mechanical energy input across a wide range of environmental conditions. This is achieved through the synergy of spontaneous capillary action and evaporation, enabling autonomous, low-intensity energy supply solutions ideal for portable devices, Internet of Things applications \cite{mandal_protein-based_2020, cheng_flexible_2018, shen_self-powered_2019} and harvesting of low-grade waste heat (below 100 $^\circ \text{C}$) produced by industrial, agricultural, and domestic processes \cite{bruckner_industrial_2015, papapetrou_industrial_2018}. When integrated with solar-driven evaporators that utilize photothermal effects \cite{ding_hybrid_2021} to significantly enhance evaporation rates, EDHV systems have the capability to also convert solar energy into electrical power. This advancement underscores their potential as a reliable and sustainable energy source. However, as demonstrations of these devices expand to encompass a broader range of materials and architectures, it is imperative to engage in the ongoing debate over the fundamental mechanisms, optimal micro- and nanostructures, and operating conditions of EDHV devices. In particular, beyond enhancing evaporation rates through photothermal effects, it is crucial to understand how external heat and light sources affect the interfacial processes essential to effective hydrovoltaic energy conversion.
\paragraph{}
In recent years, EDHV devices utilizing micro-nanoporous materials have garnered significant attention for their ability to harness energy through the movement of electrolytes within partially wetted regions \cite{fang_evaporating_2022, yoon_natural_2019, wang_ionovoltaic_2023}. This phenomenon, which occurs ahead of a liquid meniscus, facilitates the electrokinetic streaming of ions. Previous studies have identified several contributing phenomena, including the streaming potential \cite{xue_water-evaporation-induced_2017, anwar_salinity-dependent_2024, qin_constant_2020}, ionovoltaic effect \cite{yoon_natural_2019, wang_ionovoltaic_2023, jin_identification_2020}, and evaporating potential \cite{fang_evaporating_2022, yu_high_2023}. Central to these mechanisms is the critical role of directional electrolyte flow near the liquid-solid interface. Additionally, going beyond the intricate micro- and nanostructures of typical EDHV devices \cite{xue_water-evaporation-induced_2017, qin_constant_2020, yaroshchuk_evaporation-driven_2022}, we recently utilized a controlled array of silicon nanopillars to demonstrate the significant impact of the geometrical and interfacial chemical properties of the nanostructures, particularly the role of the chemical equilibrium of the surface groups in enabling high salinity operation \cite{anwar_salinity-dependent_2024}, challenging the reliance of deionized water for high performance \cite{yu_high_2023}. Furthermore, recent work introduced a passive hydrovoltaic device characterized by a limited evaporation rate and fluid permeation, capable of producing electricity through upstream proton diffusion \cite{xia_electricity_2024}. This proton diffusion arises from the chemical potential difference between the wet and dry sides of the material, resulting in sustained electricity generation owing to the gradual permeation of water. Despite these advancements, it remains unclear how the various mechanisms identified in different nanomaterials and device geometries can be effectively integrated into hydrovoltaic systems \cite{yaroshchuk_evaporation-driven_2022}, or how to design structures that harness them simultaneously. Similarly, previous studies have indicated that combining light and heat can synergistically improve hydrovoltaic performance through enhanced photothermal evaporation \cite{sun_achieving_2022, ding_hybrid_2021, gao_solar_2019}. On the other hand, in nanofluidic systems with charged interfaces, thermo-osmotic flows can convert thermal gradients into electrical currents \cite{fu_giant_2019, ouadfel_complex_2023}. However, one vital area remains largely overlooked: how photocharging and (photo)thermal effects contribute to ion migration, which is a key process for the performance of hydrovoltaic devices \cite{anwar_salinity-dependent_2024, ding_hybrid_2021}. This lack of understanding emphasizes the importance of a fundamental study of light- and heat-driven ion dynamics, potentially revealing new methods to boost hydrovoltaic efficacy. 
\paragraph{}
This work unravels the complex influence of heat and light on solid-liquid interfaces in EDHV systems, ultimately demonstrating a unified concept for EDHV architectures that transcends traditional mechanisms focused solely on ion streaming at the solid-liquid interface of the evaporating surface. By employing a top-evaporating surface and a bottom uniformly structured cm-scale silicon-dielectric (core-shell) nanopillar array (Si NPs, see Fig. \ref{fig:ch3_device_architecture_heat-light}A), we report major improvements in power output under external heating and solar illumination due to ion thermodiffusion in the electrolyte layer, as well as a combination of the photovoltaic effect and thermally enhanced surface charge at the silicon-electrolyte interface. Through a combination of experiments, numerical calculations, and theoretical modeling, we reveal a mechanism in which thermally and light-assisted ion migration plays a pivotal role in enhancing electricity generation, achieving a state-of-the-art open-circuit voltage of 1 V and an output power density of $0.25\ \text{W/m}^2$ under optimal conditions with 0.1 M concentrations. Notably, these results are achieved without the need for additional black absorbers \cite{ding_hybrid_2021, gao_solar_2019}.

\paragraph{}
Overall, our EDHV architecture introduces a completely new device concept and design strategy that leverages thermal and photovoltaic effects to enhance interfacial processes essential for hydrovoltaic energy conversion. By structurally decoupling the top evaporating surface from the bottom nanostructured layer, we establish that the system can operate in a decoupled manner. This architectural innovation represents a significant advancement in hydrovoltaic device design, allowing full exploitation of developments in solar-driven interfacial evaporation \cite{ding_hybrid_2021, tao_solar-driven_2018} while simultaneously enabling the integration of optimized hydrovoltaic components.

\section{Introduction of Key Interfacial Processes }\label{ch3_intro_interfacial_process}

Understanding the complex influence of heat and light on EDHV devices requires control over the solid-liquid interfacial properties as well as the possibility to disentangle different phenomena. In particular, as schematically represented in Fig.  \ref{fig:ch3_device_architecture_heat-light}A, light and heat inputs can modify (1) the evaporation rate at the evaporating interface, (2) the ion transport within the liquid induced by chemical potential difference, and (3) the chemical equilibrium at the solid-liquid interface, which controls the surface charge. In typical EDHV devices, these effects cannot be easily decoupled as a single material serves all these functions simultaneously. Instead, we devised an EDHV architecture where the top evaporating surface, consisting of an Ag/AgCl electrode, and the hydrovoltaic component, consisting of an array of SiNPs, are spatially decoupled by an intermediate ion-conducting layer, allowing us to address these phenomena and their interplay in a controlled manner. We will discuss this strategy with an equivalent electrical circuit in the section \ref{ch3_decoupling_strategy}. The SiNP’s solid-liquid interface, in particular, plays a key role in the system behavior and performance due to the presence of a net surface charge \cite{anwar_salinity-dependent_2024}, which can originate from both electronic and ionic contributions \cite{sun_understanding_2021, lin_quantifying_2020}. In this work, we focus on the ionic contribution arising from surface dissociation reactions. More specifically, as shown in Fig.  \ref{fig:ch3_device_architecture_heat-light}A, upon wetting, any oxide layer on the surface of the SiNPs (Fig.  \ref{fig:ch3_device_architecture_heat-light}C and \ref{fig:ch3_device_architecture_heat-light}D) will dissociate, usually resulting in a net negative surface charge $\sigma$ (orange circles). Concurrently, positive ions adsorb on the surface (pink circles), and an electrical double layer (EDL) develops within the liquid \cite{gonella_water_2021}.  This results in a voltage across the EDL in the liquid and space charge layer in silicon, as shown in Fig.  \ref{fig:ch3_device_architecture_heat-light}E.

\begin{figure}
    \centering
    \includegraphics[width=\linewidth]{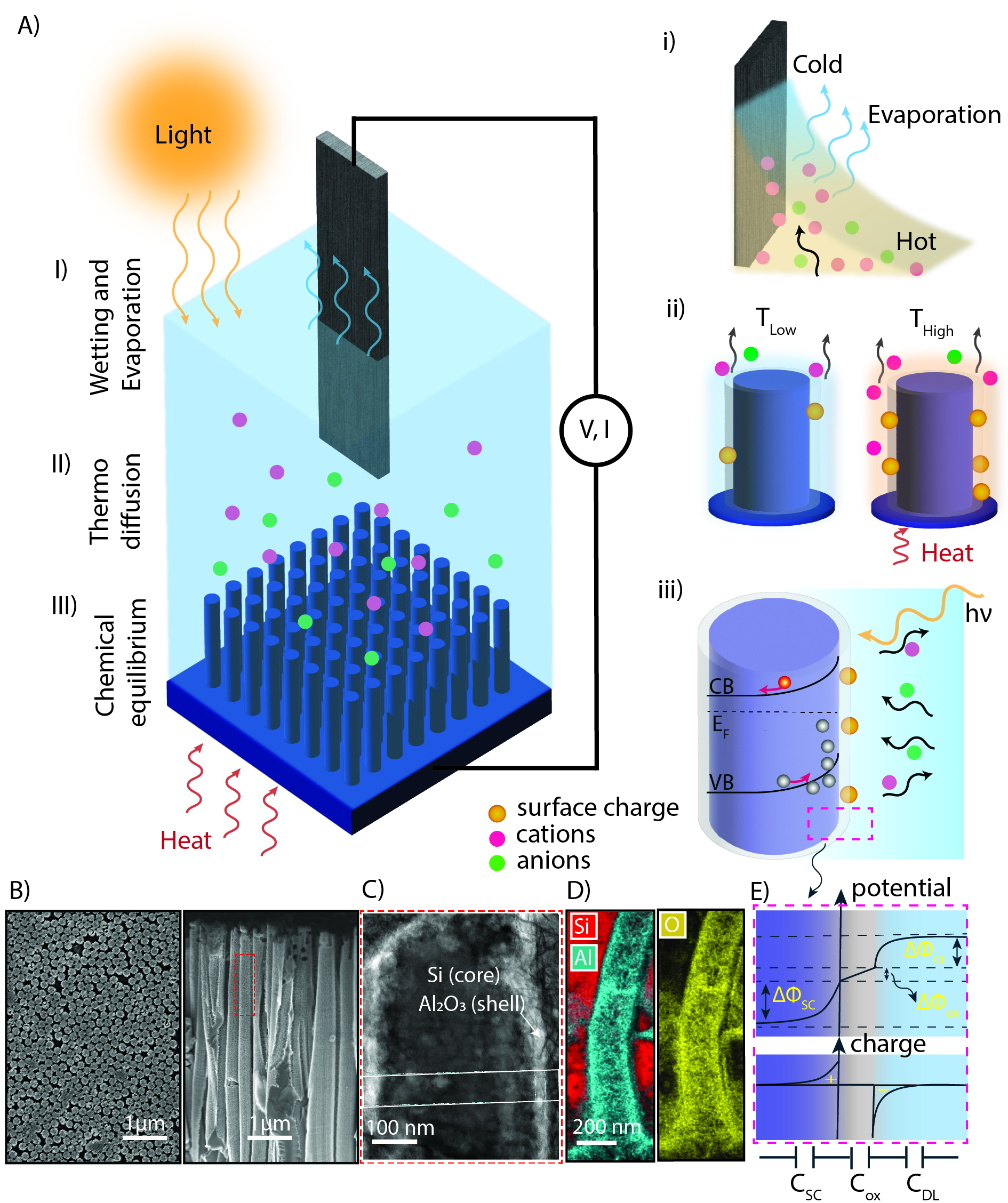}
    \captionsetup{labelfont={bf,footnotesize}}
    \caption{\footnotesize
    \textbf{Evaporation-driven Hydrovoltaic Device Architecture, Mechanisms, and Materials.} \textbf{A)} Schematic representation of the hydrovoltaic device featuring a top evaporating electrode surface and a bottom array of SiNPs immersed in water. The top and bottom components do not physically contact but are electrochemically connected through the water. The inset displays the three effects contributing to the device's performance. i) A side view of the evaporating surface with the liquid meniscus and the thermal gradient across the liquid layer. ii) An intermediate electrolyte layer and thermally tuned chemical equilibrium at the bottom nanostructures, resulting in a higher surface charge at increased temperatures. iii) photoactive nanostructure-electrolyte interface depicting the enhanced surface charge under irradiation due to electron-hole pair generation.  \textbf{B)} Scanning electron microscopy (SEM) image of the SiNPs array. (left) Top view, and (right) cross-sectional view.}
    \label{fig:ch3_device_architecture_heat-light}
\end{figure}

\begin{figure}
    \ContinuedFloat
    \captionsetup{labelfont={bf,footnotesize}}
    \caption{\footnotesize
    \textbf{(continued)} 
    \textbf{C)} Scanning transmission electron microscopy (STEM) image of a single NP. The cross-sectional cut of a single NP reveals the presence of a silicon core and \ce{Al_2O_3} shell. Intensity mapping was performed in the rectangular region. TEM-EDX image of the NP displaying the elemental maps of aluminum, silicon, and oxygen. \textbf{E)} (Top) A detailed view of the interface potential and free space charge in the silicon and electrolyte mediated by the oxide layer. The total potential is the sum of the potential in the space charge layer of silicon, the oxide layer, and the double layer of the electrolyte (bottom), as well as the free charge profile in the respective regions and the corresponding equivalent capacitance.}
\end{figure}

For oxides-aqueous interfaces, the surface dissociation can be quantitatively described using the complexation model \cite{sposito_surface_1983}, which depends on the surface oxide material, i.e., \chemfig{SiO_2}, \chemfig{Al_2O_3}, or \chemfig{TiO_2}. For example, \chemfig{TiOH} groups on the \chemfig{TiO_2} surface dissociates according to: \newline \newline
\schemestart
\chemfig{TiOH}\+ \chemfig{H_2O}
\arrow{->}
\chemfig{TiO^{-}}\+\chemfig{H_3O^{+}}
\schemestop \newline \newline
As shown above, the equilibrium reaction is governed by the equilibrium constant $K_{\text{a}}$, which exhibits temperature dependencies as given by eq. \ref{eq:ch3_equilibrium_constant_temperature_dependence}. To a first approximation, we can then recast the reaction equilibrium condition in a form that explicits the total surface charge $\sigma$ as:  
\begin{equation}\label{eq:ch3_surface_charge_light}
    \sigma = \frac{-e\Gamma}{1 + \frac{[\ce{H+}]_s}{K_a}}
\end{equation}  
where, $\Gamma$ is the density of total reactive surface sites ($\Gamma=  [\text{TiO}^-]+[\text{TiOH}]$), and $[\ce{H+}]_s$ denotes the local interfacial proton concentration. As K$_a$ increases with temperature (see methods \ref{ch3_numerical_modelling}), we can then see that the surface charge ($\sigma$) also rises with temperature. 
Importantly, eq. \ref{eq:ch3_surface_charge_light} shows that the surface charge $\sigma$ can also be varied at a constant temperature due to dynamic changes in the surface concentration of protons (\ce{[H^+]_s}). This can occur due to changes in the local pH, but interestingly, it can also happen under irradiation. The equilibration of the Fermi level \cite{mizsei_fermi-level_2002} across the solid-liquid interface, primarily driven by surface states, determines the band bending at the silicon-oxide interface (Figure \ref{fig:ch3_device_architecture_heat-light}A, panel iii). Consequently, under illumination, photogenerated charges (electrons or holes) accumulate at the interface \cite{noauthor_kinetic_nodate, mayer_photovoltage_2017}. Due to the capacitive effect of the oxide layer, a concurrent change in the surface proton concentration will occur, leading to a change in surface charge. This also highlights the importance of the oxide layer in passivating the silicon surface \cite{richter_improved_2023} and preventing any chemical reaction between silicon and water. Our photoelectrochemical test on the device indeed shows no evidence of Faradaic activity, thereby confirming a capacitive charging, rather than a Faradaic process. 
Overall, the chemical equilibrium at the interface is strongly dependent on both material properties and the temperature and ion concentration in the EDL. Light and heat triggers can have a multi-faceted influence on the interfacial chemical equilibrium, therefore affecting device behavior and performance. With our architecture, without additional black absorbers, we can decouple the effects of light and heat on the interfacial chemical equilibrium, disentangling them from a purely photothermally driven evaporation enhancement.

\subsection{Experimental Platform}
All our devices feature a bottom cm-scale regular array of SiNPs (Fig.  \ref{fig:ch3_device_architecture_heat-light}A), fabricated using a combination of Nanosphere lithography and metal-assisted chemical etching of a Si wafer (see Methods \ref{Ch3_fabrication}, Fig.  \ref{fig:ch3_device_architecture_heat-light}B). To passivate the surface and control its properties, we use atomic-layer deposition to coat the \ce{Si-SiO_2} core, where \ce{SiO_2} is the native oxide layer, with few-nm thick dielectric shells of \ce{Al_2O_3} or \ce{TiO_2} (Fig.  \ref{fig:ch3_device_architecture_heat-light}C-D and Methods \ref{Ch3_fabrication}). These two materials, in fact, have distinct chemical equilibria. In addition, as any band-bending at the silicon-oxide interface will be affected by the Fermi level of the semiconductor, and hence its doping, we specifically used three silicon dopings, Low N-doping (\( 1-20~\mathrm{\Omega.cm} \)) and high N-doping (\( <0.05~\mathrm{\Omega.cm} \)) and P-doped (\( 0.1-0.5~\mathrm{\Omega\ cm} \)), which has different space charge layer thickness and capacitance $C_{\text{sc}}$. For hydrovoltaic testing, the sample is placed inside a custom HV cell and wetted with \(250~\mathrm{\mu L}\) of deionized water containing KCl salt of varying concentrations (from \( 10~\mathrm{\mu M} \) to \( 0.1~\mathrm{M} \)). In ambient conditions (T = 22-24 $^\circ$C and humidity = 25-30\%), evaporation readily occurs. Our testing HV cell is uniquely designed to prevent contact between the solution and the bottom (Aluminum) electrode, which could lead to unwanted chemical reactions, and to ensure that only the central part of the silicon (\(1~\mathrm{cm^2}\) area) is in contact with the electrolyte. Next, the HV cell is positioned on a microbalance to track the evaporation rate. The electrical response is measured using a top Ag/AgCl placed in the liquid right above the Si NPs, and an aluminum contact previously deposited on the back surface of the Si wafer (Fig.  \ref{fig:ch3_device_architecture_heat-light}A and Methods \ref{Ch3_fabrication}). During the electrical measurements, the top electrode and bottom silicon substrate are decoupled, and the electrical circuit is complete as soon as the liquid is dispensed; thereafter, voltage and current can be measured. Heat and light stimuli are applied using a Peltier cell and a Solar simulator, respectively, while measuring the open circuit voltage, \(\mathrm{V_{oc}}\) or power output of the device under different temperatures and irradiation (see methods \ref{ch3_electrical_measurments_methods} for more details).

\section{Results and Discussion}
Due to the chemical equilibrium-controlled surface charge, at the oxide-liquid interface, an EDL is formed, and any resulting imbalance in the ion distribution along the SiNP length contributes to the measured electrical potential difference. Interestingly, we previously observed that, due to the closed bottom surface of the nanochannel, the studied geometry presents an intrinsic asymmetry in the surface-charge distribution. As a result, even in the absence of an evaporation-induced flow, a chemical potential difference ($\Phi$) exists, and therefore, a \(\mathrm{V_{oc}} \) can be measured between the top electrode and the bottom SiNPs surface \cite{anwar_salinity-dependent_2024}. $\Phi$ is quantified by eq. \ref{eq:ch3_G-C_model_surface_potential}, which depends on the temperature ($T_\text{s}$) and surface charge ($\sigma$) of the bottom silicon substrate: 

\begin{equation}\label{eq:ch3_G-C_model_surface_potential}
    \Phi = \frac{2k_B T_{\text{s}}}{e} \sinh^{-1} \left( \frac{\sigma}{\sqrt{8000 \varepsilon_0 \varepsilon_r c_0 k_B T_{\text{s}}}} \right) + \frac{\sigma}{C_{\text{Stern}}}
\end{equation}
In the following sections, we will demonstrate how the performance metrics of the EDHV devices — specifically, open-circuit voltage and power density — are influenced by changes in chemical potential under conditions of external heating and irradiation.

\subsection{Effects of Temperature Change on Chemical Equilibrium}\label{ch3_temperature_chemical_equilibrium}
\begin{figure}
    \centering
    \includegraphics[width=1\linewidth]{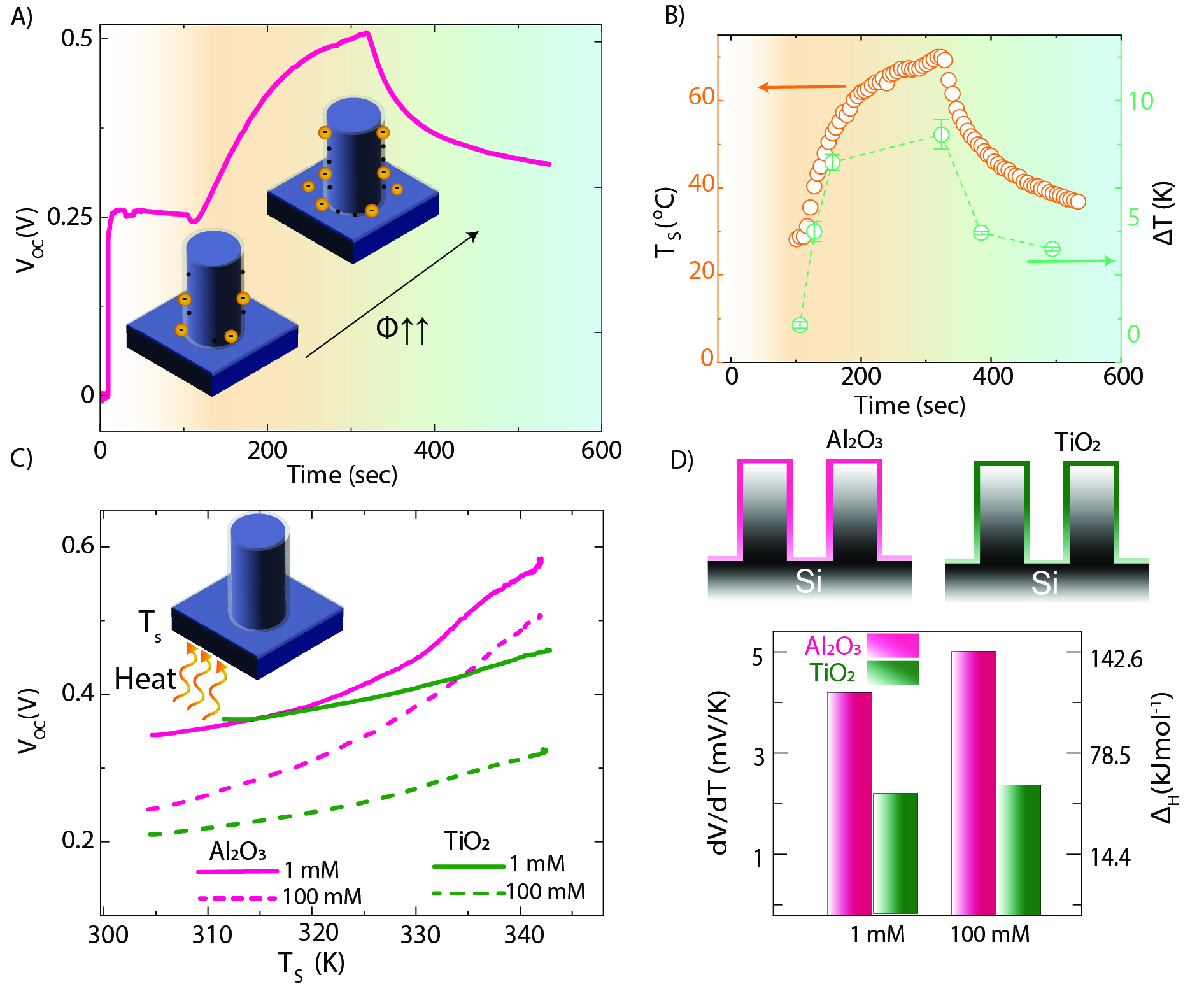}
    \captionsetup{labelfont={bf,footnotesize}}
    \caption{\footnotesize\textbf{Role of Temperature for different coatings and salinity levels.} \textbf{A)} The time trace of the measured open circuit voltage at ambient temperature is presented when the silicon surface temperature is increased and then allowed to cool down. The inset illustrates a qualitative increase in the chemical potential difference for a single NP wetted with electrolyte. \textbf{B)} The time trace of the silicon surface temperature when the heater is turned on and turned off once the maximum is reached. The right axis shows the corresponding temperature difference between the bottom silicon surface and the top electrode (at the end of the liquid meniscus). \textbf{C)} The voltage-temperature profile of the two devices was measured using the same top electrode at varying salinities. Each device consists of an identical silicon core and a dielectric shell made of \ce{Al_2O_3} and \ce{TiO_2}: 1 mM (solid lines) and 100 mM (dashed lines). \textbf{D)} (Top)Schematic representation of the core-shell nanopillars with \ce{Al_2O_3} (pink) and \ce{TiO_2} (green) shells. (bottom) The slope of the \(\mathrm{V_{oc}}\) -temperature curves for the linear regime (up to a 20 K increase in temperature) and the corresponding estimated values of $\Delta_H$.  }
    \label{fig:ch3_heat-Voc}
\end{figure}
To grasp the impact of temperature, we begin by measuring \(\mathrm{V_{oc}}\) without any heating applied. Once we switch on the heater located at the bottom of the substrate, we observe a gradual rise in temperature alongside an increase in surface charge, which leads to a notable surge in \(\mathrm{V_{oc}}\). Although an increase in temperature will lead to an increase in ionic mobility and a lower \(\mathrm{V_{oc}}\) \cite{anwar_salinity-dependent_2024}, we regard this as a secondary effect as we have always observed an increase in \(\mathrm{V_{oc}}\) with temperature for all the conditions, highlighting the surface charge increase being the dominant effect. After the temperature stabilizes, we turn off the heater and let the device cool naturally. This cooling process results in a lower temperature, which corresponds with a decrease in surface charge, thereby illustrating a gradual decline in \(\mathrm{V_{oc}}\).  

While focusing solely on the interfacing effects,  our previously validated COMSOL Multiphysics model \cite{anwar_salinity-dependent_2024}, which accounts for the liquid nanoconfinement, modified to incorporate the temperature-dependent equilibrium constants for the oxide-liquid interface (see methods \ref{ch3_numerical_modelling}), we found the electrostatic potential difference between the bulk electrolyte and bottom SiNPs to increase linearly with temperature. Therefore, we first conducted a series of experiments explicitly analyzing the open-circuit voltage generated at different temperatures of the bottom SiNPs electrode, denoted $T_{\text{s}}$. To trigger the temperature changes, we used an external Peltier heater placed beneath the SiNPs electrode. Concurrently, we monitored the temperatures using an infrared camera positioned above the sample, calibrated using a thermocouple (see methods).\\
Fig. \ref{fig:ch3_heat-Voc}A provides a time trace of the open circuit voltage of a SiNPs device with N-type low-doping silicon core nanopillars coated with \ce{Al_2O_3}. At the beginning of the experiment (dry sample), the voltage was zero. As soon as the electrolyte wets the sample, we observe a gradual increase in voltage until a steady-state value is reached (0.25V). When we activate the heater (at 125 s), the voltage experiences a noticeable rise, eventually approaching twice the initial value (0.5 V). The corresponding time-trace for $T_{\text{s}}$, as recorded by the infrared camera, is depicted in Figure \ref{fig:ch3_heat-Voc}B (empty symbols). As the heater operates, the temperature steadily climbs and ultimately stabilizes around 70 $^\circ$C for the maximum power applied to the heater. Once the heater is turned off, we observe an immediate drop in temperature and \(\mathrm{V_{oc}}\). Notably, the cooling rate is slower than the heating phase, which can be attributed to restricted pathways for heat dissipation. It is important to highlight that the temperature of the top electrode, located at the air-water interface, is observed to be slightly lower than the measured $T_{\text{s}}$ due to the cooling effect of evaporation and the differing thermal environments present at the bottom (heating source) and top (ambient air) of the setup. We quantified this temperature difference as $\Delta T$. Figure \ref{fig:ch3_heat-Voc}B illustrates the time trace of $\Delta T$ as $T_{\text{s}}$ is increased from the ambient temperature of \(\mathrm{25~ ^\circ C}\) to \(\mathrm{70~ ^\circ C}\), peaking at around 7 K. \\
We further examined the slopes of the voltage-temperature lines across various conditions, including pH, electrolyte concentration, and the initial equilibrium constant ($K_a^0$), and found minimal dependence on these external factors. Indeed, the $dV/dT$ slope is predominantly influenced by the enthalpy of dissociation of the surface groups, $\Delta_ H$, which is the chemical characteristic of the material. While this relationship is derived numerically, it underscores a significant physical dependence on the key thermodynamic parameters at play. Thus, we experimentally investigated how different surface properties affect the device's open-circuit performance metrics as a function of $T_{\text{s}}$. Fig. \ref{fig:ch3_heat-Voc}C presents the voltage-temperature profile for two distinct samples that share the same silicon N-type low-doping core but differ in their outer shells, composed of \ce{Al_2O_3} (pink curves) and \ce{TiO_2} (green curves), respectively. These experiments were conducted at two KCl concentrations: 1 mM and 100 mM. Based on the experimental results, \ce{Al_2O_3} displayed a more considerable voltage increase compared to \ce{TiO_2} for equivalent temperature rises (Fig.  \ref{fig:ch3_heat-Voc}C), and the slopes remain insensitive to changes in concentration, consistent with the simulation. We observe that the experimental curves present a linear range (approx. up to 20 K), consistent with our COMSOL model, and a super-linear regime at higher temperatures. Within the linear regime, we can extract the $dV/dT$ slope and relate it to the enthalpy of dissociation, confirming the expected trend due to the lower enthalpy of \ce{TiO_2} compared to \ce{Al_2O_3} \cite{yun_surface_2023}(Fig. \ref{fig:ch3_heat-Voc}D-E). This behavior clearly stems from the distinct chemical properties of the two materials. Furthermore, the thermal conductivity of the oxide surface, commonly low in \ce{TiO_2}—has minimal impact on the system's overall behavior.\\
As we show later, to model the experimental \(\mathrm{V_{oc}(T)}\) data beyond the linear regime, it is necessary to account for the complex interplay of the chemical equilibria and the enhancement of the evaporation rates with temperature. This is discussed in Section \ref{ch3_decoupling_strategy}, where we develop a comprehensive equivalent electrical circuit and establish an expression of \(\mathrm{V_{oc}}\) by integrating three distinct phenomena. 

\subsection{Effect of Irradiation on Chemical Equilibrium}\label{ch3_light_chemcial_equilibrium}
As described earlier, the surface charge $\sigma$ can also be varied at a constant temperature due to dynamic changes in the surface concentration of protons (\ce{[H^+]_s}) under irradiation. Thus, we conducted a comprehensive series of measurements to understand the device performance under various illumination conditions and different solid and interfacial properties. We begin by assessing the open-circuit voltage of the same samples used in the temperature-dependent study, namely N-type low-doping silicon core nanopillars coated with \ce{Al2O3} and \ce{TiO2}. The \(\mathrm{V_{oc}}\) time trace is shown in Fig.  \ref{fig:ch3_light-voc}A (green for \ce{TiO_2} and pink for \ce{Al_2O_3}), omitting the transient phase where it rises from 0 V to a steady state value for improved clarity.
Under ambient conditions, before illumination, the devices show a stable voltage (0.36\ V \ce{Al_2O_3}, 0.15\ V \ce{TiO_2}). Upon exposing the samples to solar illumination (AM 1.5, \(\mathrm{1~kWm^{-2}}\)), we observed an instantaneous increase in the measured voltage, which stabilizes at a significantly higher steady-state value (0.55 V \ce{Al_2O_3}, 0.33 V \ce{TiO_2}). Conversely, when we turned off the light source, the voltage immediately decreased, reverting to its initial value recorded in the dark. To validate the consistency of this response, we performed multiple cycles of switching the light on and off, and each cycle showed a reliable, pronounced rise and fall in voltage. We note here that due to the instantaneous nature of the \(\mathrm{V_{oc}}\) change upon illumination, photothermal effects are expected to play a minor role. Instead, photogenerated charges are contributing to the observed behavior. 
When n-doped silicon is subjected to illumination, upward band bending occurs, accumulating holes at the silicon-oxide interface \cite{mizsei_fermi-level_2002} (Fig.  \ref{fig:ch3_wavelength-intensity_voc}A). Simultaneously, on the liquid side, excess charges at the capacitive interface drive the dissociated cations of water —primarily hydronium ions (\ce{H_3O^+}) —away from the interface. It also creates an attractive force for hydroxide ions (\ce{OH^-}), drawing them toward the interface to neutralize the accumulated holes. Consequently, this process leads to a reduction in the concentration of \ce{[H^+]_s} under illumination, increasing the surface charge ($\sigma$) that in turn produces a higher \(\mathrm{V_{oc}}\) or a positive photovoltage \(\mathrm{V_{ph}}\). Moreover, a differential capacitance of the double layer, \(\mathrm{C_{DL}}\), can be defined as the change in surface charge ($\sigma$) due to a change in $\Phi $, and mathematically expressed as by $ C_{\text{DL}}=(\partial \Phi ⁄\partial \sigma )^{-1}$, which increases with surface charge. To quantify the time-dependent changes in total capacitance, which is a combination of the three capacitances as shown in Fig.  \ref{fig:ch3_device_architecture_heat-light}E, we measured the device's complex impedance in real time at a fixed frequency of 1 kHz. Fig.  \ref{fig:ch3_light-voc}A clearly shows that alterations in capacitance triggered by light irradiation, confirming the phenomenon is purely capacitive, as evidenced by the overlap between the voltage-time and capacitance-time traces during light on/off cycles. This overlap suggests that illumination promotes capacitive charging at the interface, thereby increasing the voltage.

\subsection{Photocharging of charged 
interfaces}\label{ch3_photocharging_interfaces}
\begin{figure}
    \centering
    \includegraphics[width=1\linewidth]{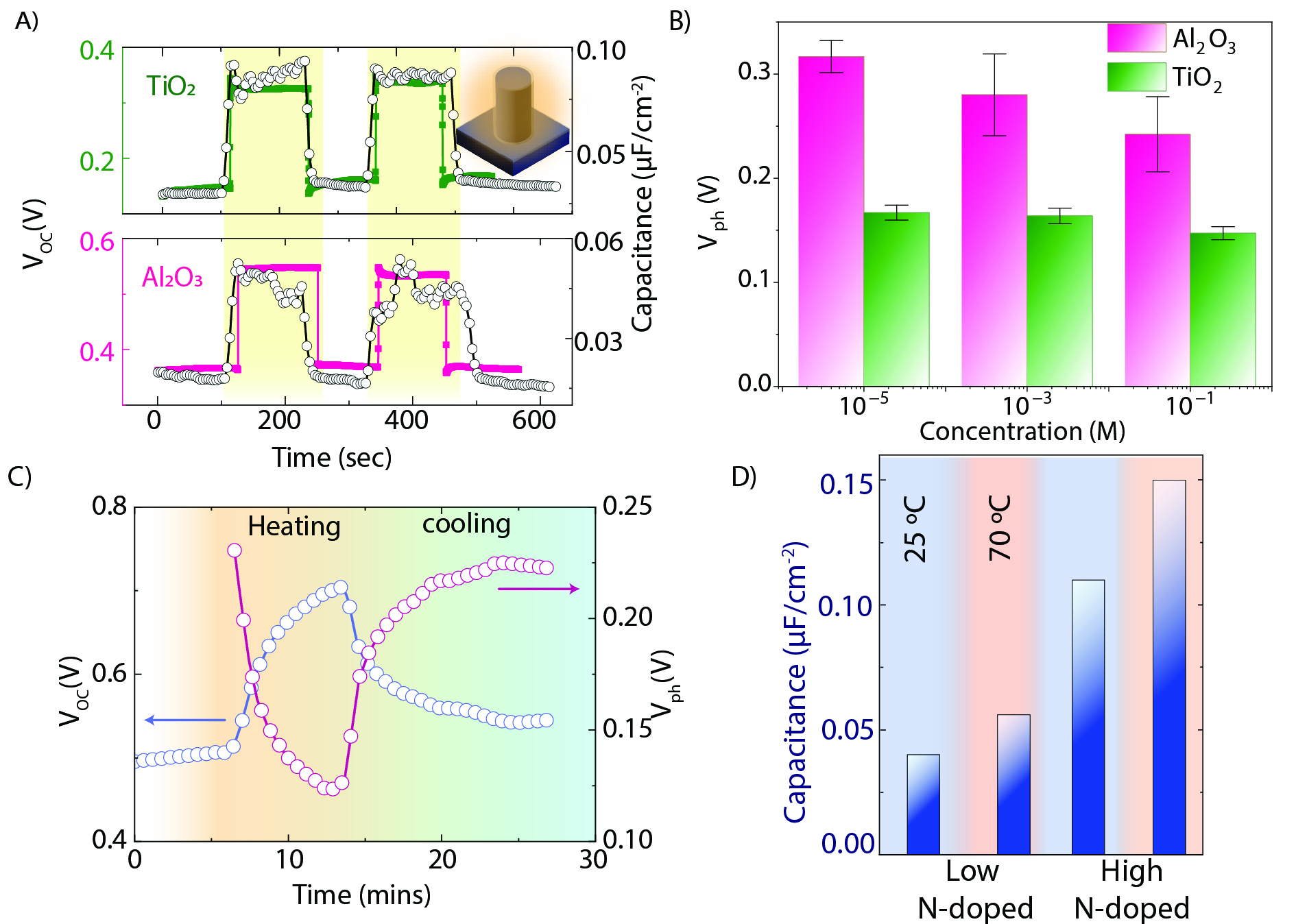}
    \captionsetup{labelfont={bf,footnotesize}}
    \caption{\footnotesize\textbf{Role of Light for different coatings and salinity levels.} \textbf{A)} Time trace of the measured open-circuit voltage and capacitance at 1 mM KCl for two devices with different dielectric shells and low N-doping (\(\mathrm{1-20~ \Omega.cm}\)). The test was conducted under ambient conditions and \(\mathrm{100~ mWcm^{-2}}\) solar illumination (yellow-shaded region). \textbf{B)} Measured photovoltage under \(\mathrm{100~ mWcm^{-2}}\) intensity for samples with \ce{Al_2O_3} and \ce{TiO_2} shells (same low N-doped silicon core) at different concentrations of KCl. \textbf{C)} Measured open circuit voltage in the dark and the corresponding photovoltage measured as the device is heated and then allowed to cool down.  \textbf{ D)} Steady-state capacitance (blue bars) values for two devices with \ce{Al_2O_3} shell, but with different doping of silicon core (low N-doped:  (\(\mathrm{1-20~ \Omega.cm}\)) and high N-doped:  (\(\mathrm{< 0.05~ \Omega .cm}\))) at 1 mM KCl. The blue and red shaded region is measured at a surface temperature of $T_{\text{s}}= T_{\text{ambient}} = 25 ^\circ \text{C}$ and $T_{\text{s}} = 70 ^\circ \text{C}$, respectively. }
    \label{fig:ch3_light-voc}
\end{figure}
We then measured the open circuit voltage of the same two low N-doping Si ((\(\mathrm{1-20~ \Omega.cm}\))) samples for different salinities (from 0.01 mM to 100 mM). Figure \ref{fig:ch3_light-voc}B presents the photovoltage, defined as $V_{\text{ph}}=V_{\text{oc}}^{\text{light}}-V_{\text{oc}}^{\text{dark}}$, for both materials across varying salinities. Our observations clearly demonstrate that the photovoltage recorded for samples with Al$_2$O$_3$ shells consistently outperformed that of the \ce{TiO_2} samples across all salinity levels tested. Our observations also reveal a significant decline in photovoltage as salinity levels increase for both sample types. Intriguingly, we further noted that High N-doped silicon samples exhibit photovoltage values that are consistently over 150 mV lower than their low N-doped counterparts. Finally, we assessed the photovoltage as a function of combined heating/cooling and irradiation (Fig. \ref{fig:ch3_light-voc}C). Interestingly, we observed that, contrary to the open circuit voltage, which increases with temperature, the photovoltage decreases with increasing temperature.\\
To better understand all of these critical findings and their relationship to the solid-liquid surface charge, we conducted capacitance measurements under all these different conditions. As shown in Fig.  \ref{fig:ch3_device_architecture_heat-light}E, the measured capacitance is linked to three capacitances in series. Firstly, the space charge layer capacitance is directly proportional to silicon doping \cite{memming_semiconductor_2015}. We thus compared the response of 2 samples with different silicon doping, specifically Low N-doping ((\(\mathrm{1-20~ \Omega.cm}\))) and high N-doping (\(\mathrm{< 0.05~ \Omega.cm}\))). As shown in Fig. \ref{fig:ch3_light-voc}D (blue, T$_{\text{s}}$ =25 $^\circ$C ) and (red, T$_{\text{s}}$ =70 $^\circ$C) shaded areas, the capacitance for the low-doped silicon sample (left two columns) was significantly lower than that of the high-doped silicon samples (right two columns), in agreement with the expected trend.  Secondly, the capacitance of the oxide layer is given by $C_{\text{ox}}=\varepsilon_0 \varepsilon_r/d $, where $\varepsilon_0$ is the permittivity of free space, $\varepsilon_r$ is the dielectric constant of the material, and d is the thickness of the oxide. The higher dark capacitance  \ce{TiO_2} compared to \ce{Al_2O_3} (Fig. \ref{fig:ch3_light-voc}A, right axis) can thus be related to the dielectric constant of anatase \ce{TiO_2} being 3-5 times higher than that of \ce{Al_2O_3} \cite{li_giant_2010}, explaining the consistently lower photovoltage for \ce{TiO_2} samples across all the salinity values (Fig. \ref{fig:ch3_light-voc}B). Thirdly, the EDL capacitance must increase with surface charge. By measuring the capacitance at different electrolyte concentrations and temperatures, we confirm that it rises with increased electrolyte concentration and surface charge. This is also in agreement with the observed trend of decreasing photovoltage with an increase in electrolyte concentration due to higher capacitance (Fig.  \ref{fig:ch3_light-voc}B) as well as with the increase in capacitance with temperature for both Low- and High N-doping of Si (Fig. \ref{fig:ch3_light-voc}D, blue bars, red shaded areas). 

\begin{figure}[h!]
    \centering
    \includegraphics[width=1\linewidth]{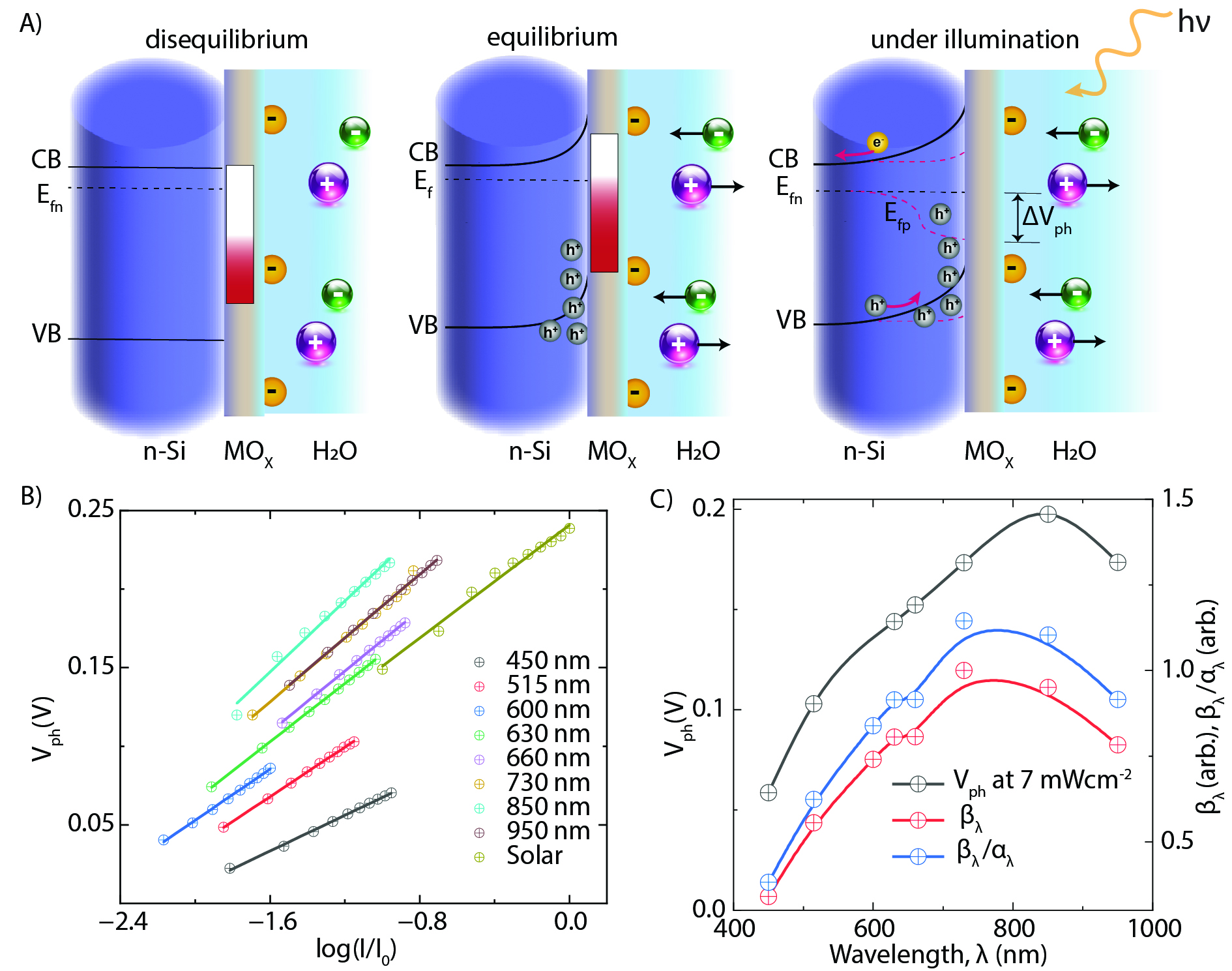}
    \captionsetup{labelfont={bf,footnotesize}}
    \caption{\footnotesize\textbf{Effect of wavelength and intensity of the irradiation.} \textbf{A)} Band diagram of the semiconductor-metal oxide-electrolyte interface. (left) The liquid-solid interface is formed, but equilibrium has not yet been reached. (middle) Equilibrium has been established, resulting in Fermi-level alignment across the interface. The red-white gradient represents the filling of surface states leading to the Fermi level pinning. (right) Under illumination, the Fermi-level splitting occurs. The difference in the quasi-Fermi levels of electrons and holes equals the measured photovoltage. \textbf{B)} Measured photovoltage for different monochromatic incidence as a function of light intensity, plotted in logarithmic scale. \ce{I_0} is equal to \(\mathrm{100~ mWcm^{-2}}\). The points corresponding to solar are the measured photovoltage for the full solar spectrum. The experimental data points are shown in circles, while the line is the linear fit. \textbf{C)} The black curve is the measured photovoltage for different at \(\mathrm{7~ mWcm^{-2}}\). The red curve is the estimated $\beta_{\lambda}$ from the linear fit. The blue curve is the value of $\beta_{\lambda}$, the normalized absorptance $\alpha_{\lambda} $ of the sample. The subscripts $\lambda $ signify wavelength-dependent physical quantities.}
    \label{fig:ch3_wavelength-intensity_voc}
\end{figure}

Thus, an increase in surface charge boosts interfacial capacitance, which influences the photovoltage, governed by the relationship between charge and capacitance. Therefore, at higher capacitance, the voltage gain tends to decrease.  This is because capacitive charging diminishes as the initial capacitance increases. Overall, these results confirm the key role of light-triggered alterations in the chemical equilibrium at the interface via capacitive photocharging, while concurrently underscoring the complex interplay among doping levels, salinity, and capacitance. \\ 
To further investigate the interplay between photocharging and interfacial processes, we finally examined how variations in light intensity influence the photovoltage response. As shown in Fig.  \ref{fig:ch3_wavelength-intensity_voc}D, due to the capacitive effect, the accumulation of holes at the silicon-oxide interface drives the dissociated cations of water, primarily hydronium ions (\ce{H_3O^+}), away from the interface. This leads to a reduction in the concentration of \ce{[H^+]_s} enabling higher surface charge ($\sigma$) that in turn produces a higher \(\mathrm{V_{oc}}\) or a positive photovoltage (\(\mathrm{V_{ph}}\)). The number of electron-hole pairs generated is directly proportional to the intensity of the light. However, not all generated carriers will lead to the charging effect as there will be a surface or bulk recombination process, which depends on the interface structure and the energy of generated carriers \cite{sze_physics_2007}. While the detailed analysis of these phenomena is beyond the scope of this manuscript, we notably observe that the number of generated charge carriers is directly proportional to the light intensity, expressed mathematically as:
\begin{equation}\label{eq:ch3_charge_carrier_intensity}
    n_h \sim \frac{\beta_{\lambda} I_{\lambda}}{E_{\lambda}}  
\end{equation}
where $\beta_{\lambda}$ is the steady-state excess carrier density generated per unit incident photon (which also needs to account for recombination rate), $I_{\lambda}$ is the incident intensity, and  $E_{\lambda}$ is the energy of the photon. On the other hand, the concentration of protons on the liquid side is given by the Boltzmann distribution \cite{memming_semiconductor_2015} as 
\begin{equation}\label{eq:ch3_proton_concetration_surface_PB_distribution}
    [H^+]_{\text{s}} \sim e^{-\frac{e \Phi}{k_B T}}
\end{equation}
which influences the chemical potential according to the relationship 
\begin{equation}\label{eq:ch3_phi-log_H}
    \Phi \sim -\log([H^+]_{\text{s}})
\end{equation}
Assuming the change in \ce{[H^+]_s} is directly proportional to the shift in \ce{n_h} we can establish the dependence of photovoltage on intensity as:
\begin{equation}\label{eq:ch3_photovoltage_intensity}
    V_{\text{ph}}\sim \Delta \Phi \sim \log \left (\frac{\beta _{\lambda}I_{\lambda}}{E_{\lambda}}\right )
\end{equation}
We conducted measurements of photovoltage across various monochromatic light wavelengths, ranging from 450 nm to 950 nm, as well as the full solar spectrum, across a broad intensity range. The results reveal that the photovoltage exhibits a clear logarithmic dependence on light intensity, with a linear trend observed when plotted on a logarithmic scale (Fig. \ref{fig:ch3_wavelength-intensity_voc}B). Additionally, we can derive estimates for the parameter $\beta_{\lambda}$ at different wavelengths, as illustrated in Fig.  \ref{fig:ch3_wavelength-intensity_voc}C, highlighting the selectivity of the wavelengths.\\Lastly, we have shown in Fig. \ref{fig:ch3_wavelength-intensity_voc}C (blue line) the value of $\beta_{\lambda}$ normalized to the absorption of the sample $\alpha_{\lambda}$ to exclude the variation in the amount of light reaching the surface for different wavelengths.

Overall, to better understand the mechanisms underlying these pronounced light-sensitive behaviors, we need to develop a more comprehensive model of the system that captures not only the solid-liquid interface but also the silicon core's light response. In the following, we introduce an equivalent circuit model and subsequently use it to gain a complete description of the device’s behavior. 

\subsection{Decoupling Strategy and Rationale for Equivalent Electrical Circuit}\label{ch3_decoupling_strategy}

\begin{SCfigure}[][h!]
    \centering
    \includegraphics[width=0.5\linewidth]{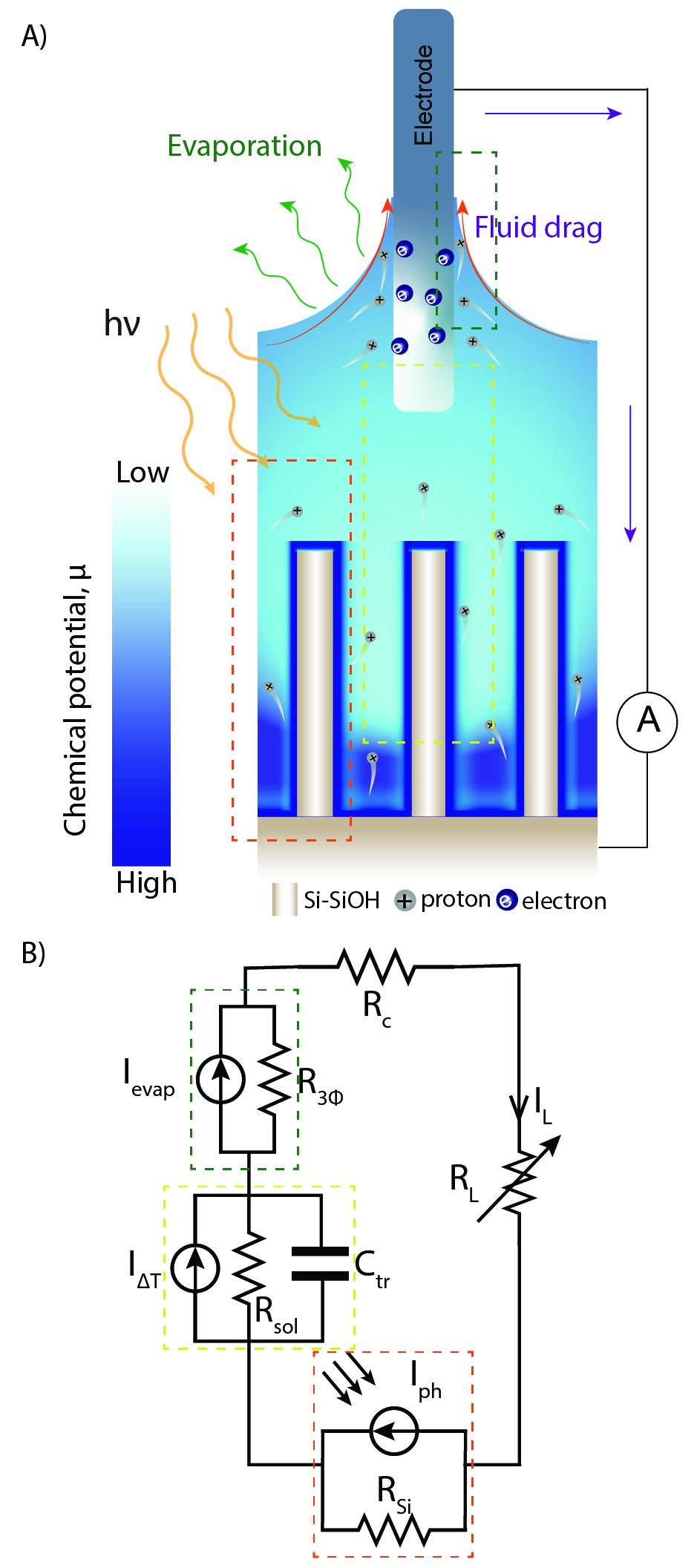}
    \captionsetup{labelfont={bf,footnotesize}}
    \caption{\footnotesize\textbf{Equivalent Circuit Model of the EDHV Device.} 
    \textbf{A)} Schematic representation of the device working principle. 
    The three regions enclosed in the box highlight the different aspects of the device that govern the electrical output. 
    The green box illustrates the generation of a potential difference across the liquid wetting front caused by evaporation-driven fluid flow along the meniscus region. 
    The yellow box illustrates the presence of a chemical potential gradient (resulting from the differences in chemical characteristics of the surface and temperature) between the wetted parts of the top and bottom electrodes. 
    The orange box shows the light-active region that generates photovoltage under solar illumination. 
    \textbf{B)} An equivalent circuit representation of the device is outlined above. 
    A contact resistance is included alongside a variable load resistance.}
    \label{fig:ch3_transfer_capacitance_equivalent-circuit}
\end{SCfigure}
Drawing from the mechanistic insights related to the interface, our goal is to enhance the understanding at the device level. We can break down our decoupled device into three distinct regions, each characterized by unique interfacial processes. As illustrated in Fig. \ref{fig:ch3_transfer_capacitance_equivalent-circuit}A, the studied system consists of three key components: the top electrode, the electrolyte layer, and the bottom nanostructured electrode, each defined by its solid/liquid interfaces. These components are crucial to the device's behavior, driven by ion streaming, chemical potential differences, and photovoltage. It is essential to recognize that the interaction between surface charges at each solid interface and the ion distribution in the electrolyte is significant; any modification to either component directly affects the overall device behavior in complex ways, establishing a non-linear dynamic feedback loop. As we outline further, the equivalent circuit depicted in Fig.  \ref{fig:ch3_transfer_capacitance_equivalent-circuit}B accurately reflects this multi-level coupling, capturing the complex behavior of the hydrovoltaic devices. 

\subsubsection{Evaporation-induced Ion Streaming at the Three-phase Boundary of the Top Electrode}
As depicted within the green rectangle, the top electrode, which serves as the positive terminal for electrical measurements, is only partially immersed in the electrolyte, with the remaining portion exposed to air. This arrangement creates a liquid meniscus and a three-phase contact line around the perimeter of the electrode. The evaporation from this region generates fluid flow along the meniscus. To assess the role of the meniscus at the top electrode, we measured the evaporation rate from our working device with a 1 cm$^2$ area SiNP bottom electrode (Low N-doping, coated with \ce{Al_2O_3}) under three different conditions: a) without any top electrode, b) with a W=0.8~mm (perimeter) top electrode, and c) with a W = 3.2~mm top electrode. We observed that the evaporation rate nearly doubled when the small electrode was introduced, compared to the scenario without an electrode. Additionally, the evaporation rate increased almost fourfold with the larger electrode, which correlates with the perimeter of the three-phase contact line. 
The flow induced by evaporation drives free ions in the liquid’s electrical double layer to stream, which can be represented by a current source, \ce{I_{evap}}, with an appropriate resistance in parallel, named $R_{3\phi}$. Equivalently, it can also be represented by a voltage source with $R_{3\phi}$ in series. Physically, the voltage source originates due to the three-phase contact boundary that generates an evaporation-rate-dependent contact potential difference between the wet and dry sides of the top electrode \cite{artemov_three-phase_2023}.
We obtained an expression for the current source \ce{I_{evap}} in the equivalent circuit (Eq. \ref{eq:ch3_evap_flux_transfer_capactiance}) that shows that it is indeed proportional to the evaporation velocity ($v_{\text{f}}$). This expression shows that \ce{I_{evap}} depends on the solvent-dependent charge flux ($\bar\sigma$) induced by the direct interaction of polar solvent molecules with the evaporating surface \cite{fang_evaporating_2022} as well as ion separation arising due to the chemical potential difference ($C_{\text{tr}} \Phi$ ) across the electrolyte. To confirm the role of polar solvent molecules in the process, we tested the device using water or ethanol under otherwise identical conditions. The measured \(\mathrm{V_{oc}}\) values for DI water and ethanol at $T_{\text{s}}$ equal concentrations \ce{25 ^$\circ$ C} are 0.552 V (0.690 V), and 0.195 V (0.336 V), respectively, confirming a clear dependence on the polar solvent molecules' interaction with the charged surfaces. These observations indicate the importance of ion streaming at the system's three-phase boundary at the top electrode and the need to include it explicitly in the equivalent circuit model. 

\subsubsection{Chemical Potential Gradient across the Electrolyte Layer}
As illustrated in the yellow rectangle, the wetted portion of the top electrode and the bottom SiNPs have different surface charges, resulting in a chemical potential difference ($\Phi $) across the electrolyte layer. Due to this, the electrolyte region exhibits capacitive effects as the surface charges of either electrode are perturbed by thermal effects or light irradiation.  We thus define a transfer capacitance $C_{\text{tr}}$ of the electrolyte, which is the capacitance arising from ion transfer across it \cite{artemov_three-phase_2023}. We have obtained an analytical expression for transfer capacitance that depends on the geometrical parameters of the bottom nanostructures as well as the size of the top electrode and the features of the related meniscus regions.
Under equilibrium conditions, the chemical potential difference would cease to exist. However, due to the temperature gradient $\Delta T$ across the electrolyte layer, the non-zero chemical potential difference persists in the steady state. Overall, the yellow region can be modeled as a current source driven by the thermodiffusion $ I_{\Delta T}$, in parallel with a resistor, which accounts for the ionic resistance of the electrolyte solution $R_{\text{sol}}$, and a capacitor $C_{\text{tr}}$. 

\subsubsection{Photovoltage at the Silicon-Oxide-Water Interface}
 
To account for the influence of light, we need to examine changes in interfacial band bending at the silicon-oxide-water interface as the sample is illuminated. As illustrated in Fig. \ref{fig:ch3_wavelength-intensity_voc}A (right panel), the generation of electron-hole pairs creates a quasi-Fermi level splitting that results in a photovoltage (\(\mathrm{V_{ph}}\)), as shown in Fig.  \ref{fig:ch3_light-voc}C. As highlighted in section \ref{ch3_photocharging_interfaces}, critical factors such as doping, salinity, and the oxide layer intricately influence the capacitive photocharging mechanism, which, in turn, profoundly affects the magnitude of V$_{\text{ph}}$. Therefore, the strength and effectiveness of the photocharging capacitive component are fundamentally encapsulated in the equivalent circuit by the value of V\textsubscript{ph}, underscoring its crucial role in this phenomenon \cite{memming_semiconductor_2015, monch_semiconductor_2001}. This can be represented by a voltage source \cite{li_kinetic_2021, monch_semiconductor_2001} or, equivalently, by a current source \(\mathrm{I_{ph}}\), in parallel with a resistance of the photoactive interfacial region (\(\mathrm{R_{Si}}\)). 
Based on the described equivalent electrical circuit, we can derive a few critical equations concerning the device response (full details in \ref{ch3_analytical_modelling} ). These can be used to quantitatively analyze the experimental data and clarify the response of this general model of an EDHV system. Remarkably, the experimental non-linear  \(\mathrm{V_{oc}}\) trend with temperature (Fig. \ref{fig:ch3_heat-Voc}B) can be finally elucidated thanks to the obtained expression for \(\mathrm{V_{oc}}\) (eq. \ref{eq:ch3_VOC_expression}, see \ref{ch3_analytical_modelling}), which includes the three contributions discussed in section \ref{ch3_intro_interfacial_process}: (i) evaporation-induced ion streaming, (ii) the chemical potential difference, and (iii) photovoltage. 

\begin{equation}\label{eq:ch3_VOC_expression}
    V_{\text{oc}}=v_{\text{f}} \bar \sigma  r_{3 \phi } + \Phi  (1+v_{\text{f}} r_{3 \phi} C_{\text{tr}} ) + V_{\text{ph}}
\end{equation}
 where  $\bar \sigma = \rho_{\text{f}} L_{3 \phi}$ self-charge density, which is not the bound surface charge, but instead free space charge per unit volume $\rho_{\text{f}}$, multiplied by the length of the wetted portion of the electrode exposed to air, denoted by $L_{3\phi}$, $v_{\text{f}}$ is equal to the evaporative mass flux divided by the density of water. Notably, both $\Phi$ and $v_{\text{f}}$ increase with temperature, so their product in the second term yields the observed quadratic dependence of the open circuit voltage on temperature.

\section{Power Output analysis using an equivalent electrical circuit}

\begin{figure}[htbp]
    \centering
    \includegraphics[width=\linewidth]{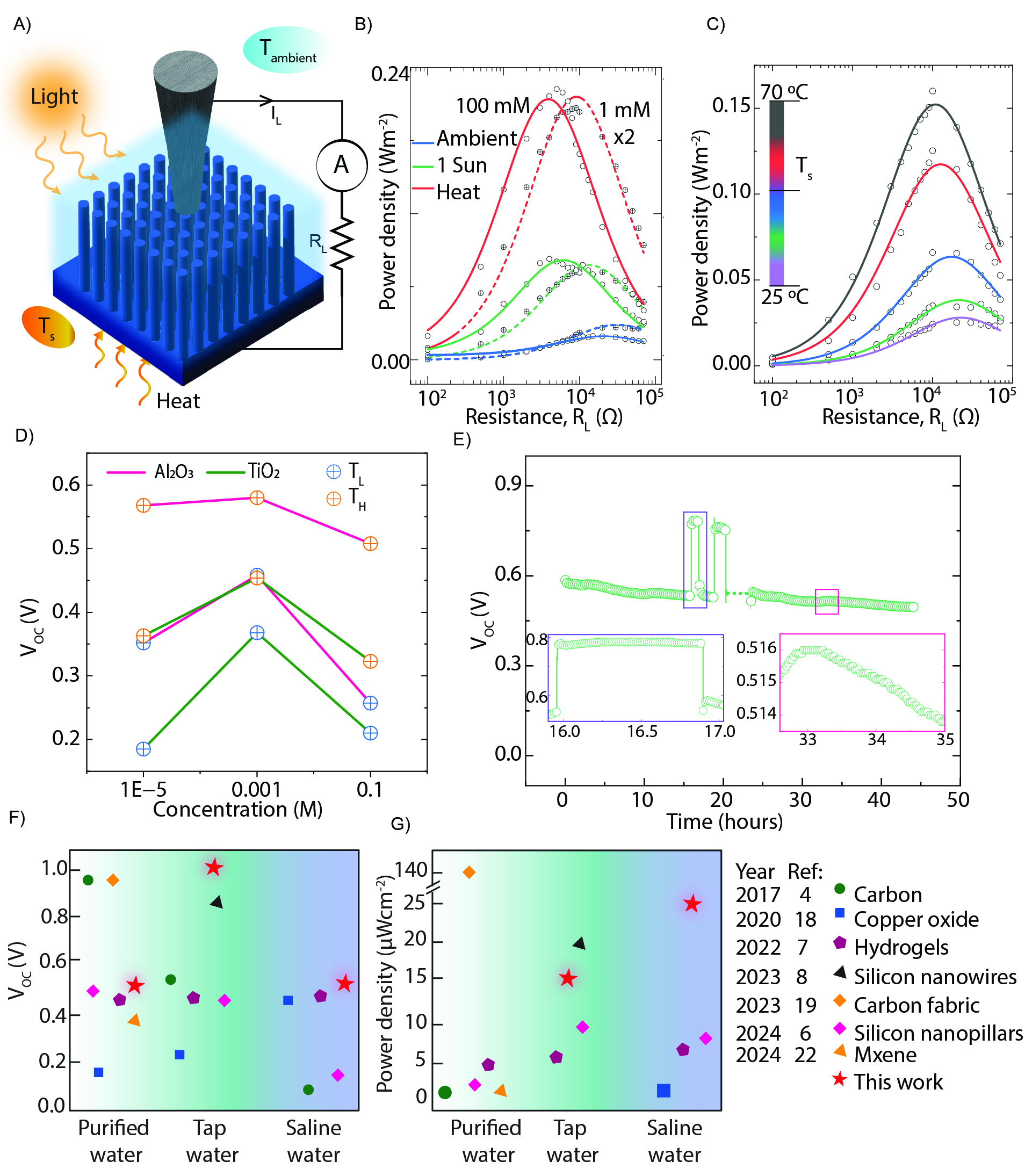}
    \captionsetup{labelfont={bf,footnotesize}}
    \caption{\footnotesize
    \textbf{Device performance as a function of materials, electrolyte, and environmental conditions (heat/light).} 
    \textbf{A)} Schematic representation of the device measurement configuration with various external loads for different surface temperatures and under illumination. The measurements were conducted at ambient temperature, with or without solar illumination, or at elevated surface temperatures. 
    \textbf{B)} Power–load resistance profiles for various conditions (ambient: \(\mathrm{T_s=25~^\circ C;~1~Sun:~1~kWm^{-2}}\) solar intensity and \(\mathrm{T_s=25~^\circ C;~heat~T_s=70~^\circ C}\)) at 1 mM and 100 mM KCl. The profiles demonstrate an increase in peak height and a shift of the peak toward lower resistances under illumination and at elevated temperatures. The empty (crossed) circles represent experimental values, while the solid (dashed) lines illustrate fits based on our model for 1 mM (100 mM) KCl. 
    \textbf{C)} Power–load resistance profiles for various surface temperatures (\(T_{\text{s}}\)) at 1 mM KCl demonstrate an increase in peak height and a shift in the peak toward lower resistance at elevated temperatures. The empty circles represent experimental values, while the solid lines depict the fit to our two-parameter family of curves. \textbf{D)} Steady-state open-circuit voltage values for two devices with \ce{Al_2O_3} and \ce{TiO_2} dielectric shells but with the same silicon core (low N-doped: \(\mathrm{1 - 20~\Omega\,cm}\)) at different salinity values. The blue and orange circles correspond to measurements at \(T_{\text{s}}=T_{\text{ambient}}=25~^\circ C\) and \(T_{\text{s}}=T_{\text{H}}=70~^\circ C\), respectively. \textbf{E)} Long-term stability testing of the device. The open-circuit voltage was measured continuously for 45 hours using an Ag/AgCl electrode. The sample was illuminated (solar spectrum, \(\mathrm{100~mWcm^{-2}}\)) in 1-hour on/off cycles for $\approx$15 hours total.}
    \label{fig:ch3_power_curves_heat-light}
\end{figure}

\begin{figure}[h!]
    \ContinuedFloat
    \captionsetup{labelfont={bf,footnotesize}}
    \caption{\footnotesize
    \textbf{(continued)} 
     The inset provides enlarged views of the regions marked in purple and pink rectangles: the purple region shows an enlarged light off-on-off cycle, clearly illustrating the instantaneous \(\mathrm{V_{oc}}\) response to illumination (photocharging, not photothermal). The red region shows stability with < 2 mV change over 2 hours. 
    \textbf{F)} Open-circuit voltage and 
    \textbf{G)} power-density comparison at different salt concentrations for various EDHV devices.}
\end{figure}
Finally, we investigate the influence of various operational parameters on the power output of our device, emphasizing the effects of temperature, solar illumination, and electrolyte concentration. Fig.  \ref{fig:ch3_power_curves_heat-light}A shows a schematic representation of power curve measurements by sweeping load resistances in the range \(\mathrm{100~ \Omega - 70~k\Omega}\). Firstly, the peak value (position) of the power curves increases (shift towards lower resistance) as the environmental conditions change from ambient (\(\mathrm{T_s=25 ^\circ C}\), no light irradiation) to 1 Sun (\(\mathrm{T_s=25 ^\circ C \ and \ 1~kWm^{-2}}\) light irradiation) and then heat (\(\mathrm{T_s=70 ^\circ C}\), no light irradiation), as shown in Fig. \ref{fig:ch3_power_curves_heat-light}B.  Notably, the measured power curves indicate that increasing the concentration of KCl from 1 mM to 100 mM nearly doubles the maximum power output despite a lower open circuit voltage at higher concentrations.  Furthermore, as observed in the power curves at varying temperatures (Fig. \ref{fig:ch3_power_curves_heat-light}C), the maximum power at \(\mathrm{T_s=70 ^\circ C}\) surpasses the ambient temperature output by over five times, demonstrating the profound impact of thermal conditions on device performance. Moreover, the parabolic dependence of power output and voltage across the load resistance underscores the importance of optimizing both aspects to achieve high performance. 
The power output is governed by the equivalent electrical circuit, which allows us to derive an expression that captures the relationship between the load resistance and output power. The simplification of this expression into a two-parameter family of curves
\begin{equation}\label{eq:ch3_power_curve_parametric}
    P=\frac{A(T,I,c_0 ) R_L}{[B(T,I,c_0 )+R_L ]^2}  
\end{equation}
demonstrates the distinct roles of parameters A and B, which depend on temperature, electrolyte concentration, and solar intensity. By fitting this expression to the experimental data in Fig. \ref{fig:ch3_power_curves_heat-light}B-C, we have compiled the values of the A and B parameters for a range of different conditions, which can be used to obtain the current and voltage utilized to obtain the power density. The peak power output, represented by  $P=A⁄4B$, occurs at a specific load resistance $ R_L=B$, indicating that careful tuning of these parameters is crucial. Notably, achieving double the maximum power output at 100 mM compared to 1 mM KCl concentration—despite a lower open circuit voltage—highlights the significant impact of internal resistance on peak power output, as indicated by parameter B in our study model. In fact, the increase in the open circuit voltage with temperature reinforces the need to consider material choices carefully; a transition from \ce{TiO_2} to \ce{Al_2O_3} yielded remarkable enhancements of 90\% and 57\% in \(\mathrm{V_{oc}}\) at \(\mathrm{T_s=25 ^\circ C}\) and \(\mathrm{T_s=70 ^\circ C}\), respectively (Fig. \ref{fig:ch3_power_curves_heat-light}D). 

Furthermore, the influence of silicon doping illustrates another pathway for enhancing power output. High N-doped \ce{Al_2O_3} samples displayed a 28\% increase in open circuit voltage over low N-doped counterparts, leading to a 1.6-fold improvement in power output. Lastly, long-term stability tests revealed that the device consistently maintained a steady voltage output for over 50 hours. Remarkably, this stable performance persisted even under continuous light exposure, as demonstrated in Fig.  \ref{fig:ch3_power_curves_heat-light}E, underscoring the robustness and reliability of the EDHV architecture. This finding lays the groundwork for future research in material engineering to optimize device performance across diverse environmental conditions. In summary, the interplay between concentration and material properties offers a rich avenue for improving power output in EDHV devices under external heating and solar illumination.\newline
The experimental results, consistent with the proposed model, demonstrate significant enhancements in power output resulting from the combined influence of heat and light. This improvement is not solely due to increased evaporation but is predominantly driven by improved ion dynamics, caused by light-induced photocharging at the silicon-oxide interface and by a thermally enhanced chemical potential at the oxide-electrolyte interface. Thus, it provides a unified perspective that incorporates thermal and photovoltaic effects, in addition to the electrical potential generated at the top evaporating surface.  

\section{Conclusion}
This work establishes a spatially and functionally decoupled hydrovoltaic platform that disentangles the key physical processes—evaporation, ion transport, and interfacial chemical equilibrium—that govern energy generation in evaporation-driven water-based systems. By strategically introducing an intermediate electrolyte layer, we enabled independent yet coupled control over each of these phenomena, laying the foundation for a unified, generalizable framework for designing hydrovoltaic operation beyond material-specific constraints. By systematically modulating experimental parameters such as doping concentration, oxide composition, and salinity, we demonstrate predictable, tunable performance under illumination via photocharging at charged interfaces. This approach transforms empirical observations into physically grounded design principles. We have developed a robust semi-quantitative predictive model based on an equivalent electrical circuit, emphasizing transfer capacitance as a central element. This model not only captures the interdependence between the physical structure and interfacial charge dynamics, but also provides explicit analytical expressions that directly link these phenomena to device output. Our study is the first to explicitly investigate the coupled influence of light and heat on solid-liquid interfacial equilibria in hydrovoltaic systems. We reveal that capacitive photocharging and thermally driven modulation of surface ion equilibria are the dominant mechanisms underpinning energy conversion when interfacial environments are carefully engineered. Without relying on photothermal evaporation enhancement, our approach achieves state-of-the-art hydrovoltaic performance across a wide salinity range (Fig. \ref{fig:ch3_power_curves_heat-light}F), underscoring the robustness and scalability of the design. Together, these insights provide a rigorous physical foundation for advancing hydrovoltaic technologies and offer a roadmap for designing next-generation devices capable of harvesting ubiquitous solar and low-grade thermal energy sustainably.

\renewcommand{\thesection}{M\arabic{section}}

\section*{Methods}

\renewcommand{\thefigure}{M\arabic{figure}}
\renewcommand{\thetable}{M\arabic{table}}
\renewcommand{\theequation}{M\arabic{equation}}
\renewcommand{\thesubsection}{M\arabic{subsection}}
\setcounter{figure}{0}
\setcounter{table}{0}
\setcounter{equation}{0}

\subsection{Fabrication of core-shell Silicon nanopillar array}\label{Ch3_fabrication}
Metal-assisted chemical etching (MACE) of crystalline silicon, combined with nanosphere lithography, was used to fabricate a cm-scale array of SiNPs \cite{wendisch_large-scale_2020-1, anwar_salinity-dependent_2024}. It involves the self-assembly of polystyrene (PS) nanospheres at the water-air interface. Then, the non-closed-packed assembly of PS nanospheres was compressed to a pressure of approximately $25-30\ \text{N/m}^2)$ using the Langmuir-Blodgett system, which resulted in a homogeneous closed-packed hexagonal lattice of PS nanospheres \cite{thangamuthu_reliable_2020}. The closed-packed monolayer was transferred to Piranha-cleaned Silicon (P-type $0.1-0.5 \ \Omega.\text{cm}$, and N-type $<0.005$ and $1-20 \ \Omega.\text{cm}$) substrates diced into $2\text{cm} \times 2\text{cm}$ chips. Plasma etching was used to reduce the diameter of the PS nanospheres, which had an initial diameter of d = 400 nm. After the gold-sputtering deposition of a thickness of 20 nm and lift-off, a gold nanomesh is formed and used as an etching mask for MACE. Before gold sputtering, a 2-3 nm of Ti is sputtered as an adhesion layer. This forms a stable contact between gold and the substrate, preventing delamination during MACE. The liftoff was done by putting the substrate in Toluene and ultrasonicating it at moderate power for 3-5 minutes at room temperature. Finally, MACE was performed by placing the substrate in an aqueous \chemfig{HF}/\chemfig{H_2O_2} solution containing 10\% and 2\% HF and \chemfig{H_2O_2}, respectively. The diameter of the NPs is controlled by changing the time of plasma trimming of the polystyrene nanosphere, while the length of the NPs is controlled by changing the MACE time. The fabricated SiNPs were then coated with a dielectric shell of \chemfig{Al_2O_3} and \chemfig{TiO_2} via atomic layer deposition, using trimethylaluminum (TMAl) and titanium tetrachloride (\chemfig{TiCl_4}) as aluminum and titanium precursors, respectively. The obtained sample was then treated with oxygen plasma (60 seconds, 1000 W) to enhance its hydrophilicity.

\subsection{Electrical \& Thermal Measurements}\label{ch3_electrical_measurments_methods}
Fig.  \ref{fig:ch3_device_architecture_heat-light}A and \ref{fig:ch3_power_curves_heat-light}A illustrate the configuration for electrical measurements, where SiNPs serve as the active substrate. The Al layer functions as the back electrode, while Ag/AgCl (unless stated otherwise) is employed as the top electrode. We used an Ag/AgCl electrode, as it is considered fully reversible, which ensures that the electrode entirely consumes the charges accumulated in its EDL at overpotential, which are practically zero. There is no unwanted potential difference, which induces a conduction current. The open-circuit voltage and complex impedance measurements were performed using a CHI bipotentiostat. The impedance measurements were performed at a zero applied DC voltage with an amplitude of 5 mV at a fixed frequency of 1 kHz. The resistance was added to the circuit to measure voltage and power at different load resistances. The bipotentiostat was connected in series or parallel for current or voltage measurements, respectively.  For temperature-dependent measurements, the device's temperature was varied by changing the voltage applied to the Peltier element, ranging from an ambient temperature of 25 $^\circ$C to 70 $^\circ$C. Temperature is measured using an Infrared camera (Fortric 600s) with 0.2 frames per second. A type-K thermocouple was used for calibration.  A solar simulator (Newport 66 984-300XF-R1 Xe lamp) with an AM 1.5 G filter was used as the light source for the light-dependent study. For a wavelength- and intensity-dependent study, we used the Ossila LED solar simulator (G2009A1). The intensity of the monochromatic light was measured with a detector (Newport, 818-UV/DB) and a handheld laser power meter (Newport, 843-R-USB). The electrical measurements were performed under conditions similar to those described above for ambient conditions. After each measurement, the samples were carefully rinsed with Isopropanol, followed by DI water. Then, the sample was placed in a DI water bath (maintained at 60°C) for 5-10 minutes with continuous stirring at 500 RPM. Then, finally, the sample was placed in the oven, which was maintained at 80 $^\circ$C. Furthermore, to minimize the effect of salt precipitation on the output, the sweep of concentration was always performed starting from DI water, followed by low concentration, and then higher concentrations. Therefore, for the present study, we expect the effect of salt precipitation at high concentrations to be minimal. We have verified the measurement repeatability at 2 M and 4 M KCl concentrations. 

\subsection{Deriving an analytical expression for the performance metrics}\label{ch3_analytical_modelling}
The total space charge density per unit area of the meniscus region is the sum of the self-charge density $\bar \sigma = \rho_{\text{f}} L_{3 \phi}$ (when the electrode was inserted in a bulk electrolyte) and the charge density arising due to the intermediate electrolyte layer, which is equal to transfer capacitance times the chemical potential difference ($C_{\text{tr}} \Phi$). Note that $\bar\sigma$  is not the bound surface charge, but instead free space charge per unit volume $\rho_{\text{f}}$, multiplied by the length of the wetted portion of the electrode exposed to air, denoted by $L_{3\phi}$. The term \ce{I_{Evap}} is equal to total charge times $v_{\text{f}}$, which is equal to the evaporative mass flux divided by the density of water.
\begin{equation}\label{eq:ch3_evap_flux_transfer_capactiance}
I_{\text{Evap}} = v_{\text{f }}\left( \rho_{\text{f}} L_{3\phi} W + C_{\text{tr}} \Phi W \right)
= W v_{\text{f}} \left( \bar{\sigma} + C_{\text{tr}} \Phi \right)
\end{equation}
The current driven through the load resistance is obtained by solving the equivalent electrical circuit. The resistive element $R_{3\phi} = \rho_{3\phi}  L_{3 \phi}/W = r_{3\phi}/W$, where $L_{3 \phi} $and W are the length and perimeter of the meniscus region. So, we use $r_{3\phi}$ instead, as it is independent of the top electrode size. 
\begin{equation}\label{eq:ch3_load_current_resitance}
I_L = \frac{I_{{\Delta} T} R_{\text{sol}} (1 + v_{\text{f}} r_{3\phi} C_{\text{tr}}) + v_{\text{f}} \bar{\sigma} r_{3\phi} + V_{\text{ph}}}{R_{\text{Si}} + R_{\text{sol}} (1 + v_{\text{f}} r_{3\phi} C_{\text{tr}}) + r_{3\phi}/W + R_C + R_L}
\end{equation}
The open-circuit voltage is obtained under the conditions, $I_L R_L (R_L \rightarrow \infty)$, and under no current through the external load, we have the relation satisfying $I_{\Delta T} R_{\text{sol}} = \Phi$. Thus, 
\begin{equation}\label{eq:ch3_VOC_expression_1}
V_{\text{oc}} = \Phi \left( 1 + v_{\text{f}} r_{3\phi} C_{\text{tr}} \right) + v_{\text{f}} \bar{\sigma} r_{3\phi} + V_{\text{ph}}
\end{equation}
Using Ohm’s law, we obtained the power consumed by the resistive load, and it is given by: 
\begin{equation}\label{eq:ch3_power_load_resitance}
P_L = \frac{\left[I_{\Delta T} R_{\text{sol}} (1 + v_{\text{f}} r_{3\phi} C_{\text{tr}}) + v_{\text{f}} \bar{\sigma} r_{3\phi} + V_{\text{ph}}\right]^2 R_L}{\left [R_{\text{Si}} + R_{\text{sol}} (1 + v_{\text{f}} r_{3\phi} C_{\text{tr}}) + r_{3\phi}/W + R_C + R_L\right]^2} = \frac{A(T,I,c_0)R_L}{[B(T,I,c_0)+R_L]^2}
\end{equation}
By incorporating the condition of maximum power $d(I_L^2 R_L)/(dR_L )=0$, we obtained the value of load resistance and voltage in terms of the functions A and B, as follows:
\begin{subequations}\label{eq:ch3_performance_metric}
\begin{align}
R_L  &= B(T,I,c_0 ) \label{eq:1a} \\
 V &= \frac{\sqrt{A(T,I,c_0)}}{2} \label{eq:1b} \\
P_{max} &= \frac{A(T,I,c_0 )}{4B(T,I,c_0)}\label{eq:1c}
\end{align}
\end{subequations}
\subsection{Numerical calculation to obtain the voltage-temperature dependence}\label{ch3_numerical_modelling}

We developed a 3D numerical model (COMSOL) to solve the Nernst-Planck-Poisson equations and determine the equilibrium distributions of ions and the resulting electrostatic potentials. For modeling, we considered different modes of ion transport using the Nernst-Planck equation for the transport of dilute species coupled with the Poisson-Boltzmann equation for the equilibrium distribution of ions. To perform these calculations, however, we must first identify an equivalent simplified geometry using the approach described in our prior work \cite{anwar_salinity-dependent_2024}. The vertical cross-sectional cut plane shows the electric field and the distribution of ionic species concentrations for a particular geometric size and a 0.1 mM bulk ionic concentration. We used surface charge density as the boundary condition, which is not a fixed value but is instead governed by the equilibrium reaction \ref{eq:ch3_equilibrium_reaction_general}, in which the protonated metal oxides dissociate as follows:
\begin{equation}\label{eq:ch3_equilibrium_reaction_general}
    MOH + H_2O 	\rightleftharpoons MO^- + H_3O^+
\end{equation}

The equilibrium constant for this reaction is given by Eq. \ref{eq:ch3_equilibrium_constant_mass_action_law}
\begin{equation}\label{eq:ch3_equilibrium_constant_mass_action_law}
    K_a = \frac{[H_3O^+][MO^-]}{[MOH][H_2O]}
\end{equation}
This results in a variable surface charge density that depends on the surface potential at the Stern plane through the surface concentration of protons ([H$^+$]$_{\text{s}}$) \cite{behrens_charge_2001} given by Eq. \ref{eq:ch3_surface_charge_light}.
which is temperature-dependent due to the temperature dependence of the equilibrium constant \cite{vordonis_effect_nodate}. The temperature dependence is governed by Eq. \ref{eq:ch3_equilibrium_constant_temperature_dependence}:
\begin{equation}\label{eq:ch3_equilibrium_constant_temperature_dependence}
K_a = K_{a}^0 \exp \left[ -\frac{\Delta _H}{R} \left( \frac{1}{T} - \frac{1}{T_0} \right) \right]
\end{equation}
The material’s chemical characteristic $\Delta_H$ is the enthalpy of dissociation of the surface groups. Then, we swept $\Delta_H$ and obtained voltage-temperature lines for various conditions. This enabled us to obtain the open circuit voltage (surface charge) as a function of temperature for multiple conditions, such as by varying electrolyte concentration, pH, and K$_a^0$. Finally, we constructed a correlation between the slopes of voltage-temperature lines and their values $\Delta _H$. The material’s chemical characteristic $\Delta _H$, the enthalpy of dissociation of the surface groups, is a thermodynamic property related by the Vant-Hoff equation. This enabled us to obtain the open-circuit voltage as a function of temperature under various conditions, such as by varying electrolyte concentration, pH, and $K_a^0$. Then, we swept $\Delta_H$ and obtained voltage-temperature lines for various conditions. Finally, we constructed the correlation between the slopes of voltage-temperature lines and the value of $\Delta_H$ .

\section*{Acknowledgments}
T.A. and G.T. acknowledge the support of the Swiss National Science Foundation (Starting Grant 211695 and Korean-Swiss Science and Technology Cooperation Fund IZKSZ2$-$188341). T.A. also acknowledges the support of the Swiss Government Excellence fellowship. The authors acknowledge the support of Milad Sabzehparvar for his assistance with Scanning Electron Microscopy imaging and of Vasily Artemov for insightful discussions during manuscript preparation. The authors also acknowledge the support of the following experimental facilities at EPFL: the Center for MicroNanoTechnology (CMi) and the Interdisciplinary Centre for Electron Microscopy (CIME).
\section*{Author contributions}
G.T. and T.A. conceived the study. T.A. conducted the experiments and developed the models under the supervision of G.T. All authors contributed to the writing and revision of the manuscript. 
\bibliography{references}

\end{document}